\documentclass[letterpaper,11pt]{article}
\usepackage{natbib}
\usepackage{tabularx} 
\usepackage{amsmath}  
\usepackage{graphicx} 
\usepackage[margin=1in,letterpaper]{geometry} 
\usepackage[final]{hyperref} 
\usepackage{authblk} 
\hypersetup{
	colorlinks=true,       
	linkcolor=blue,        
	citecolor=blue,        
	filecolor=magenta,     
	urlcolor=blue         
}

\begin{document}

\title{An Approach to the Quantum Hall Effect in Three- Dimensional Electron Systems}

\author{M. A. Hidalgo

miguel.hidalgo@uah.es}

\affil{Departamento de Física y Matemáticas, University of Alcalá, Spain}

\maketitle

\begin{abstract}
We present a theoretical framework to describe the integer quantum Hall effect (IQHE) in three-dimensional (3D) electron systems. This extends our previous single-electron approach, which was successfully applied to two-dimensional (2D) systems such as semiconductor quantum wells and graphene, where it provided insights into both the IQHE and the fractional quantum Hall effect (FQHE). Starting from the graphene model—where the unconventional sequence of Hall plateaux, $2(2n+1)$, naturally emerges from Landau quantization—we generalize the formulation to 3D semimetals with low carrier density and high mobility, where recent experiments have reported signatures of the QHE. The density of states under a magnetic field is derived using the Poisson summation method, incorporating Gaussian Landau-level broadening, spin splitting, and thermal damping. In graphene, this reproduces both Shubnikov–de Haas oscillations and quantized Hall conductivities, consistent with gate-voltage and magnetic-field dependence. For 3D systems, the model accounts for strong band anisotropy by introducing an effective correction to the cyclotron frequency, and also by considering large effective $g$-factors, as observed in materials such as ZrTe$_5$, HfTe$_5$, and Cd$_3$As$_2$. From the calculated density of states and carrier concentration, we derive semiclassical expressions for the diagonal and Hall conductivities. The resulting Hall conductivity exhibits quantized values proportional to $2e^2/(h\lambda_F)$, where $\lambda_F$ is the Fermi wavelength, in agreement with theoretical predictions and experimental observations of 3D quantum Hall states. Simulated magnetotransport curves reproduce both Hall plateaux and Shubnikov–de Haas oscillations under realistic parameter sets. Our results demonstrate that the IQHE in 3D semimetals can be understood as a natural extension of the single-electron Landau quantization framework originally developed for 2D systems. This provides a unified picture of quantum magnetotransport across dimensions, highlighting the crucial role of low carrier density and high mobility. The model further suggests new avenues for analyzing thermodynamic and transport properties in 3D systems under quantum Hall conditions.

\end{abstract}

\section{Introduction}
The integer quantum Hall effect (IQHE), discovered in two-dimensional electron gases (2DEGs) by Klaus von Klitzing in 1980 \cite{klitzing1980new}, marked a milestone in condensed matter physics. The quantization of the transverse (Hall) conductivity at low temperatures and under strong magnetic fields, expressed in integer multiples of the conductivity unit $(e^2/h)$, was found to be remarkably robust against impurities and disorder, enabling its use in redefining fundamental physical constants. At the same intervals of magnetic field or gate voltage where Hall plateaux are observed, minima—or even zeros—appear in the diagonal conductivity, i.e., the Shubnikov–de Haas (SdH) effect. The discovery of the IQHE was soon followed by the observation of the fractional quantum Hall effect (FQHE) by \cite{tsui1982two}, where quantization occurs at fractional values of $(e^2/h)$.

While the IQHE can be observed in any two-dimensional electron system (2DES) confined in quantum wells (QWs), including MOSFETs and semiconductor heterostructures, the FQHE requires high-mobility heterostructures such as GaAs–AlGaAs. Several decades later, both the IQHE and FQHE, were also observed in graphene, \cite{novoselov2004electric}, \cite{castro2009electronic}, \cite{goerbig2011electronic}.

Despite their similarities, the integer quantum Hall effects in QWs and graphene show important differences. In conventional QW semiconductors, the IQHE plateaux appear at integer values $n = 1,2,3,\dots$ (or at $2n$ if spin degeneracy is not lifted). In contrast, in monolayer graphene (MLG), the plateaux follow the sequence $\pm 2, \pm 6, \pm 10, \pm 14,\dots$, i.e., $\pm 2(2n+1)$.

The recurrence of these sequences across different 2DESs underscores the fundamental nature of both the IQHE and the FQHE, while also highlighting their close relationship: whenever FQHE states are observed, integer plateaux also appear.

From a theoretical standpoint, the IQHE is widely considered as a manifestation of electron localization in the quantum limit. A major conceptual advance came from the topological perspective: Thouless, Kohmoto, Nightingale, and den Nijs (TKNN) showed that the integer Hall conductance in a 2D periodic system is a topological invariant—the Chern number—explaining its remarkable precision and robustness against disorder,\cite{thouless1982quantized}, \cite{haldane1988model}. By contrast, the FQHE is considered to be driven by strong electron–electron correlations in partially filled Landau levels. Laughlin’s proposal of a correlated, incompressible quantum liquid, based on trial wavefunctions, provided a theoretical foundation for the FQHE,\cite{laughlin1983anomalous}.

Thus, while the IQHE and FQHE remain deeply connected, they are described within different theoretical frameworks: electron localization and topological invariants in the first case, correlated quantum liquids in the second. However, a more recently alternative perspective is offered by Hidalgo’s approach, \cite{hidalgo2021quantum}, which unifies the integer and fractional QHE in QWs and graphene within a single framework. This model emphasizes that robust Landau quantization and plateau formation can emerge from fundamental single-particle physics and can reproduce experimental features as functions of both magnetic field and gate voltage.

Traditionally, the QHE has been regarded as a purely two-dimensional phenomenon, since quantization arises from discrete Landau levels and edge states in 2D systems. A long-standing question, however, is whether intrinsically three-dimensional (3D) systems can host quantized Hall responses, and under what mechanisms. Halperin first proposed in 1987, \cite{halperin1987possible}, that quantized Hall conductance in 3D could arise from electronic instabilities such as charge-density waves (CDWs), which open gaps in the energy spectrum. Subsequent theoretical work identified several possible routes to a “3D QHE”: (i) stacking weakly coupled 2D quantum Hall layers, (ii) field-induced instabilities of the 3D Fermi surface that create density modulations, and (iii) topological mechanisms involving Fermi-arc surface states and Weyl orbits in semimetals, \cite{xiao2022integer}. Other studies showed that electron–phonon interactions, rather than electron–electron interactions, could induce the required CDW at experimentally accessible fields, \cite{Fenton1987}.

Despite decades of proposals, clear experimental evidence for a 3D QHE was only reported in 2019, in the Dirac semimetal ZrTe$_5$, \cite{tang2019three}, \cite{galeski2021origin}. In this system, experiments observed near-zero longitudinal resistivity together with quantized Hall plateaux whose values depended on the Fermi wavelength along the field direction. This breakthrough was attributed to a magnetically induced CDW that split the 3D system into effectively decoupled 2D quantum Hall layers, enabling quantized Hall plateaux and dissipationless longitudinal transport.

Topological semimetals, such as Cd$_3$As$_2$ and nodal-ring semimetals, have since emerged as promising platforms for realizing the 3D QHE. In these systems, topologically protected surface states—such as Fermi arcs—can form closed cyclotron orbits through tunneling between Weyl nodes, effectively mimicking a 2D electron gas and supporting quantized Hall conductance, \cite{Potter2014}, \cite{zhang2019quantum}, \cite{moll2016transport}. Quantized transport consistent with QHE-like plateaux has also been reported in nanostructures and thin films of 3D Dirac/Weyl materials (e.g., Cd$_3$As$_2$), where Weyl-orbit physics produces thickness-dependent quantization and surface-dominated transport, \cite{Xiu2015}.

Although the microscopic origins vary across materials and remain under debate, a common feature of all systems showing 3D QHE signatures is their semimetallic nature. Low carrier density, small Fermi surfaces, and high mobility allow these materials to reach the extreme quantum limit (with only the lowest Landau level occupied) under experimentally accessible magnetic fields.

The aim of the present work is to provide an alternative framework for analyzing the QHE in 3D electron systems, extending our model previously developed for the QHE and SdH oscillations in 2DESs, particularly in graphene \cite{hidalgo2014study}, \cite{hidalgo2021quantum}. As experimental evidence highlights the importance of semimetals, we begin in Section 2 by summarizing our approach for graphene—our starting point thanks to its semimetallic character. Section 3 addresses the 3D case: Subsection 3.1 derives the density of states in the presence of a magnetic field, a key quantity for magnetotransport, while Subsection 3.2 presents the diagonal and Hall conductivities. Section 4 concludes with a summary and outlook.

\section{Quantum Hall and Shubnikov-de Haas effects in graphene}

Graphene, a single layer of carbon atoms arranged in a honeycomb lattice, exhibits a unique electronic structure characterized by linear band dispersion and gapless Dirac cones at the $K$ and $K'$ points, where the Fermi level lies at the intersection of the valence and conduction bands. Its vanishing density of states at the charge-neutrality point, together with exceptionally high carrier mobility, makes graphene an ideal two-dimensional counterpart to low-carrier-density semimetals. These features give rise to an unconventional IQHE, distinct from that observed in quantum wells. Remarkably, the IQHE in graphene persists even at room temperature, with Hall plateaux appearing at $n = \pm 2, \pm 6, \pm 10, \dots$.

Because of its semimetallic character, graphene provides a natural starting point for our approach to understanding the phenomenon in three-dimensional electron systems. In our model, the characteristic plateau sequence $2(2n+1)$ emerges directly from Landau quantization, \cite{hidalgo2014study}. The formulation remains within a single-electron framework, as originally developed for semiconductor QWs, yet successfully reproduces the experimentally observed dependence of both diagonal and Hall conductivities on gate voltage and magnetic field.

\subsection{Density of states in graphene under the application of a magnetic field}

In this subsection, we summarize the basic results for graphene that are necessary to understand the extension we present for three-dimensional electron systems.

The six corners of the first Brillouin zone fall into two groups, yielding two inequivalent points—the Dirac points (DPs), $K$ and $K'$. Since each DP hosts two symmetric bands, we assume identical effective masses for the valence and conduction bands, i.e, $m^*_{VB}$=$m^{*}_{CB}=m^{*}$, where $m^*_{VB}$ and $m^*_{CB}$  denote the effective masses of the valence and conduction bands, respectively.

To determine the quantized electron states under an applied magnetic field $B$, we construct the Hilbert space for the dynamical states at each DP as the tensor product of the dynamical spaces of the two bands, $\varepsilon^B=\varepsilon^B_{VB} \otimes \varepsilon^B_{CB}$, where $\varepsilon^B_{VB}$ and $\varepsilon^B_{CB}$ correspond to the valence and conduction bands, respectively. (These dynamical states are equivalent at both $K$ and $K'$.) Because both bands contribute under the magnetic field, the system acquires two degrees of freedom, and the energy spectrum is naturally described by that of an isotropic two-dimensional harmonic oscillator.

Thus, the quantized energy levels in graphene, for each Dirac point $K$ and $K'$, take the form

\begin{equation}
    E_n=(i+j)\hbar \omega_c=n\hbar \omega_c=nE_0
\end{equation}

with \emph{i}, \emph{j}, natural numbers related to each dynamical space, \emph{n}=0, 1, 2…, $\omega_c$ being the cyclotron angular frequency 

\begin{equation}
    \omega_c=\frac{eB}{m*}
\end{equation}

Here, $m^*$ is the effective mass of the electrons. In Eq. (1), $E_0 = \hbar \omega_c$ is the Landau-level spacing.

This two-dimensional harmonic oscillator structure, combined with spin and valley degeneracies, naturally leads to the unconventional Hall plateau sequence $\pm 2(2n+1)$ observed in graphene, distinguishing it from the integer sequence of conventional quantum wells.

From Eq. (1), the density of states (DOS) for each Dirac point can be obtained using the Poisson summation formula. This approach allows us to replace the discrete sum over Landau levels with an expression that separates the smooth background term, the density of states at zero magnetic field, from the oscillatory contributions responsible for quantum oscillations. Accordingly, the DOS for each Dirac point is written as

\begin{equation}
g^{DP}(E)=g_0[1+2\sum_{p=1}^{\infty}A_{\Gamma,p}A_{S,p}cos\left( 2\pi p\eta \right)]
\end{equation}

$\emph{p}$ is the summation index corresponding to the $\emph{p}$ harmonic, and $g_0=\frac{m^*}{\pi\hbar ^2}$ the two dimensional density of states in absence of magnetic field for each Dirac point, (K and K'). In equation (3), we have

\begin{equation}
     \eta=\frac{E}{\hbar \omega_c}
\end{equation}

and 

\begin{equation}
A_{\Gamma,p}=e^{\left( -\frac{1}{2}\frac{p^2\Gamma^2}{\hbar^2\omega_c^2} \right)}
\end{equation}

is the Landau-level broadening term that accounts for the interaction of electrons with defects and impurities. For simplicity, we assume a Gaussian shape for each energy level, with a width $\Gamma$ that is independent of the magnetic field, 

The additional term in the sum, $A_{S,p}$, corresponds to the Zeeman factor, which accounts for the effect of the magnetic field on the electron spin and has the expression

\begin{equation}
A_{S,p}=cos(p \pi \frac{g^*m^*}{2 m_0})
\end{equation}

Here, $g^*$ is the generalized (effective) gyromagnetic factor, and $m_0$ is the free electron mass. 

The next step is to consider the twofold degeneracy of the Dirac points, $K$ and $K'$, whose spectra are assumed to be identical. Hence, the total density of states of electrons in graphene under an applied magnetic field is given by the linear superposition of the contributions from each Dirac point:

\begin{equation}
g^{total}(E)=g^K(E)+g^{K'}(E)
\end{equation}

Once $g^{\rm total}(E)$ is obtained from Eq. (3) for each Dirac point, the corresponding electron density can be determined as

\begin{equation}
n^{DP}=\int_{0}^{E_F}g^{DP}(E)f^0(E)dE=n_0+\frac{2eB}{h}\sum_{p=1}^{\infty}\frac{1}{\pi p}A_{\Gamma,p}A_{S,p}A_{T,p}sin\left( 2\pi p\eta_F \right)]=n_0+\delta n\end{equation}

Here, $E_F$ is the Fermi energy, and $f^0(E)$ the Fermi–Dirac distribution function. $n_0$ denotes the electron density at zero magnetic field (or zero gate voltage) for each Dirac point, while $\delta n$ represents the variation in electron density arising from the quantized density of states. The effect of temperature, which determines the occupation of each state, is incorporated through the thermal damping term $A_{T,p}$.

\begin{equation}
A_{T,p}\ = \frac{z}{sinh(z)}
\end{equation}

with

\begin{equation}
z=2\pi p\frac{kT}{\hbar\omega _c}  
\end{equation}

where \emph{k} is the Boltzmann constant and \emph{T} the temperature.

Therefore, from equation (8), we can determine the total carrier density in graphene:

\begin{equation}
n^{total}=n^K+n^{K'}
\end{equation}

On the other hand, at low temperatures, the electron population near the Fermi level can be approximated as
$f(E)\sim \Theta(E_F-E)$ where $\Theta$ is the Heaviside step function and $E_F$ the Fermi energy. Hence, we can consider that this electron population close to the Fermi level can be determined by

\begin{equation}
N^{DP}(E_F))\sim n_0^{DP}\frac{g^{DP}(E_F)}{g_0}     
\end{equation}

where $g^{DP}(E_F)$ is given by equation (3) at the Fermi level with 

\begin{equation}
     \eta_F=\frac{E_F}{\hbar \omega_c}
\end{equation}

A more detailed derivation of all this expression can be found in \cite{hidalgo2014study} and \cite{hidalgo2021quantum}.

\subsection{Model for magnetoconductivities in graphene under quantum Hall conditions}

The diagonal and Hall conductivities for each Dirac point can be calculated using the semiclassical formulas, taking into consideration of the quantized density of states and electron density. Then we have for the diagonal conductivity:

\begin{equation}
\sigma_{xx}^{DP}=\frac{\sigma_{xx}^0}{(1+\omega_c^2\tau^2)}  
\end{equation}

where $\tau$ is the electron relaxation time and $\omega_c$ the cyclotron frequency, equation (2). Since the main contribution to the diagonal term comes from electrons near the Fermi level, the term in the numerator of Eq. (14) can be simplified using equation (12) as

\begin{equation}
\sigma_{xx}^0=\frac{e^2\tau}{m^*}N^{DP}\left( E_F \right)=\frac{e^2\tau}{m^*}n_0\frac{g^{DP}\left( E_F \right)}{g_0}   
\end{equation}

On the other hand, concerning the Hall magnetoconductivity, $\sigma_{xy}$, the semiclassical expression is given by

\begin{equation}
   \sigma_{xy}^{DP}=-\sigma_{xy}^0\frac{\omega_c\tau}{1+\omega_c^2\tau^2}
\end{equation}

That at high magnetic field will be simplified to

\begin{equation}
   \sigma_{xy} ^0=\frac{e^2\tau}{m^*}n^{DP}
\end{equation}

Therefore, we can determine both magnetoconductivities in graphene:

\begin{equation}
   \sigma_{xx}^{{total}}=\sigma_{xx}^{{K}}+\sigma_{xx}^{{K'}}
\end{equation}

\begin{equation}
   \sigma_{xy}^{{total}}=\sigma_{xy}^{{K}}+\sigma_{xy}^{{K'}}
\end{equation}

From both components of the magnetoconductivity tensor, Eqs. (18) and (19), we can directly obtain the elements of the magnetoresistivity tensor through the equations

\begin{equation}
    \rho_{xx}=\frac{\sigma_{xx}}{\sigma_{xx}^2+\sigma_{xy}^2}
\end{equation}

and

\begin{equation}
    \rho_{xy}=-\frac{\sigma_{xy}}{\sigma_{xx}^2+\sigma_{xy}^2}
\end{equation}

These expressions relate the measurable resistivities to the semiclassical conductivities and fully describe the magnetotransport behavior of graphene under quantum Hall conditions.

In Figures 1 and 2, we present the results of magnetotransport in graphene as obtained from our model. The values of the physical parameters used in the simulations are detailed in the captions of each figure. Figure 1 shows the simulations as a function of gate voltage, including the Hall magnetoconductivity (panel a) and both magnetoresistivities: the diagonal (Shubnikov–de Haas) and the Hall components. Figure 2 presents the same quantities as in Figure 1, but now as a function of the applied magnetic field.

\section{Quantum Hall and Shubnikov-de Haas effects in a 3D electron systems}

So far, we have noted that the phenomenon of the quantum Hall effect (QHE) in three-dimensional electron systems has been observed in semimetallic materials: 1) Bulk 3D semimetals: prominent example is the observation of a three-dimensional QHE in bulk ZrTe$_5$, a layered topological semimetal with very low carrier density \cite{tang2019three}, \cite{galeski2021origin}, and similarly in HfTe$_5$, \cite{Wang2020}; 2) topological semimetal nanostructures: For instance, Cd$_3$As$_2$, where experiments on thin films and microstructures revealed quantum oscillations and quantized Hall plateaux attributed to Weyl orbits \cite{Xiu2015}; 3) multilayer or engineered thin-film systems: Examples include WTe$_2$, InAs, and other tailored multilayer structures designed to exhibit quantum Hall phenomena under controlled conditions.

\subsection{Density of states of a 3D-dimensional electron system under the application of a magnetic field}

Our starting point for developing the 3D model is the previously established approach for the quantum Hall effect (both integer and fractional) in graphene.

Although many 3D semimetals are often described using linear band approximations, to be consistent with the three-dimensional character and to reflect the real possibility of observing the QHE in 3D electron systems, we begin by considering the density of states of a 3D electron gas at zero magnetic field. This is given by the standard expression:

\begin{equation}
    g_0^{3D}\left( E \right)=\frac{\sqrt{\left( 2m^* \right)^3}}{4\pi^2\hbar ^3}\sqrt{E}
\end{equation}

where $m^*$ is the effective mass of the electrons. This serves as the baseline DOS, which will later be modified to include Landau quantization, level broadening, and other magnetic-field effects relevant for 3D QHE phenomena.

The first step is to determine the energy levels for each electron in the 3D electron system now under the application of an external magnetic field, and once the Landau-type energy levels are established, the corresponding density of states at finite magnetic field can be derived.

As in the case of graphene (Section 2), the quantized energy spectrum for a 3D electron system under a magnetic field we assume can be described using an isotropic two-dimensional harmonic oscillator, as in Eq. (1). (Unlike graphene, in the present 3D system we do not consider the existence of two distinct electron pockets.)
Therefore, under the application of a magnetic field, the density of states of electrons can be expressed as

\begin{equation}
g^{3D}(E)=g_0^{3D}[1+2\sum_{p=1}^{\infty}A_{\Gamma,p}A_{S,p}cos\left( 2\pi p\eta \right)]
\end{equation}

where the argument of the cosine is

\begin{equation}
     \eta=\frac{E}{\hbar \omega_c}
\end{equation}

We suppose a Gaussian broadening of the Landau levels to account for the interactions of electrons with impurities and defects in the 3D system. This broadening is taken to be independent of the magnetic field. For the terms associated with Landau-level broadening and spin splitting, we use the same expressions as for the 2D electron gas, given in Eqs. (5) and (6). 

Once the density of states is established, we can proceed to determine the electron density in the 3D electron gas, including the density at the Fermi level. That is:

\begin{equation}
   n=\int_{0}^{E_F}g^{3D}\left( E\right)f^0\left( E \right)dE=n_0+\delta n 
\end{equation}

where a thermal damping additional term just like equations (9) and (10) appear. 
On the other hand, at low temperatures, and as a first approach we assume a density of electrons at the Fermi level given by

\begin{equation}
   N(E_F)=\int_{E_F-\frac{k_BT}{2}}^{\infty }g^{3D}\left( E\right)f^0\left( E \right)dE\sim E_Fg^{3D}\left( E_F \right)
\end{equation}

where the Fermi level is now determined from the electron density $n_0$ as

\begin{equation}
    E_F=\frac{\left( 4\pi^2 \right)^\frac{2}{3}\hbar ^2}{2m^*}n_0^{^\frac{2}{3}}
\end{equation}

However, two important points must be discussed: the generalized gyromagnetic factor, $g^*$, and the effective mass, $m^*$. In these 3D semimetallic materials, due to band anisotropy and topology, both quantities can be complex.

In particular, the gyromagnetic factor can reach values around 50 \cite{feng2015large, wang2016gate, sun2020large, wang2021thermodynamically}.

On the other hand, the cyclotron frequency, $\omega_c$ [Eq. (2)], depends on the effective mass. To account for this, we introduce a correction factor, $\alpha$, which effectively captures the influence of the complex, band-dependent effective mass on the orbital dynamics of electrons under a magnetic field. This factor is treated as a parameter to be determined by comparing simulations with experimental measurements. Specifically, in Eqs. (23) and (25) we use a modified cyclotron frequency, $\alpha \omega_c$.

Regarding the effective mass, we adopt values on the order of 0.04 $m_0$, as reported in the literature \cite{kamm1987fermi, pariari2015probing, yuan2016observation}.

\subsection{Model for the magnetoconductivity tensor for the 3D-dimensional electron systems under quantum Hall conditions}

As in the 2D case, we use semiclassical expressions for both conductivities. Accordingly, the diagonal conductivity for the 3D electron system can be written as

\begin{equation}
\sigma_{xx}=\frac{\sigma_{xx}^0}{(1+\omega_c^2\tau^2)}  
\end{equation}

where now

\begin{equation}
\sigma_{xx} ^0=\frac{e^2\tau}{m^*}N\left( E_F \right)=\frac{e^2\tau}{m^*}n_0\frac{g^{3D}\left( E_F \right)}{g_0^{3D}}   
\end{equation}

Concerning on the Hall magnetoconductivity we have

\begin{equation}
   \sigma_{xy}=-\sigma_{xy}^0\frac{\omega_c\tau}{1+\omega_c^2\tau^2}
\end{equation}

That at high magnetic field will be simplified to

\begin{equation}
   \sigma_{xy} ^0=\frac{e^2\tau}{m^*}n
\end{equation}

An important magnitude in the phenomenon of the Hall effect in 3D electron systems is the fundamental Hall plateau, given by, \cite{halperin1987possible},

\begin{equation}
    \sigma_{xy}^H=\frac{2e^2}{h \lambda_F}
\end{equation}

where $\lambda_F$ is the wave length of electrons at the Fermi level, obtained from equation (27):

\begin{equation}
    \lambda_F=\frac{2\pi}{\sqrt[3]{3\pi^2n_0}}
\end{equation}

From both components of the magnetoconductivity tensor, we can directly obtain the elements of the magnetoresistivity tensor using the same expressions given in Eqs. (20) and (21).

In Figures 3 and 4, we present the results of magnetotransport in 3D semimetallic material as obtained from our model. The values of the physical parameters used in the simulations are detailed in the captions of each figure. Figure 3 shows the simulations as a function of the magnetic field of the Hall magnetoconductivity (panel a) and both magnetoresistivities: the diagonal (Shubnikov–de Haas) and the Hall components (panel b). Figure 4 presents the same quantities as in Figure 3 but for other physical parameters of the 3D electron system.

\section{Discussion and Conclusion}

In this work, we have presented a unified theoretical framework for describing the integer quantum Hall effect (IQHE) in both two- and three-dimensional electron systems. Our approach builds on the single-electron Landau quantization formalism, which successfully reproduces the magnetotransport properties of two-dimensional electron systems (2DESs), including both the integer and fractional quantum Hall effects (FQHE) in semiconductor quantum wells (QWs) and monolayer graphene. In particular, for graphene, the unconventional Hall plateau sequence 2(2n+1) emerges naturally from the model, reproducing the experimentally observed gate-voltage and magnetic-field dependence of both diagonal and Hall conductivities (see Figures 1 and 2).

The model’s success in 2D systems demonstrates that even complex phenomena such as the FQHE can be captured using a first-principles single-electron framework, without invoking strong electron correlations, when appropriate considerations—such as Landau-level broadening, spin splitting, and thermal damping—are included, \cite{hidalgo2021quantum}. This unification provides a conceptual bridge linking integer and fractional quantum Hall phenomena in a single theoretical picture.

Extending this framework to three-dimensional (3D) electron systems, we have shown that the IQHE can arise naturally in semimetals with low carrier density and high mobility, such as $ZrTe_5$, \cite{tang2019three}, \cite{galeski2021origin}, $HfTe_5$, \cite{wang2020approaching}, and $Cd_3As_2$, \cite{Xiu2015}. By generalizing the density of states and including a key material-specific parameter: an effective mass correction, $\alpha$, to the cyclotron frequency, the model reproduces the formation of Hall plateaux and Shubnikov–de Haas oscillations in 3D semimetals (see Figures 3 and 4). Our simulations are in agreement with recent experimental observations of 3D quantum Hall states and suggest that the quantization of Hall conductance in 3D systems scales with the Fermi wavelength along the magnetic field direction.

The results emphasize the crucial role of low carrier density, high mobility, and band anisotropy in enabling the 3D quantum Hall effect. In particular, the introduction of an effective cyclotron frequency $\alpha\omega_c$ captures the influence of complex effective mass tensors, while high $g^*$ values reflect strong spin splitting typical of these semimetals. 

Importantly, our approach remains grounded in single-particle physics, showing that many features of the 3D QHE can be understood without invoking strong electron correlations.

In summary, our theoretical framework provides a unified description of quantum Hall phenomena across dimensions. It reproduces experimental observations in both 2D and 3D systems and offers predictive power for exploring magnetotransport in novel semimetals. Furthermore, following  our  view implemented in the 2D electron systems, \cite{hidalgo2023equilibrium}, the presented approach opens avenues for studying thermodynamic properties in 3D quantum Hall systems, bridging the gap between two- and three-dimensional topological electron physics in all their aspects.


\begin{figure}
    \centering
    \includegraphics[width=0.5\linewidth]{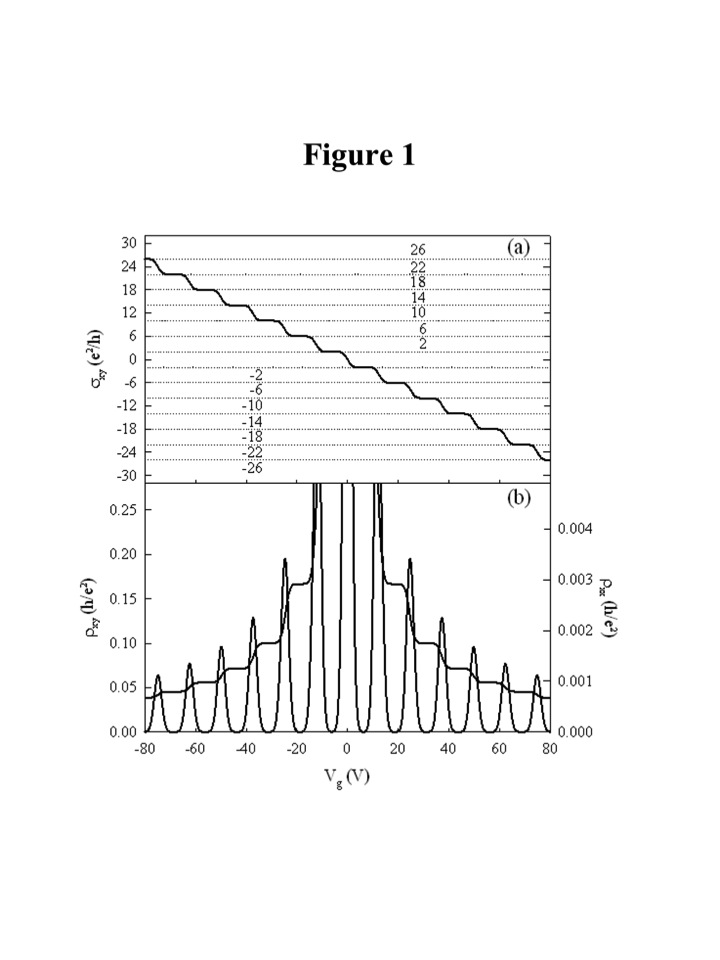}
    \caption{Simulation of the Hall magnetoconductivity (panel a) and the Hall and diagonal magnetoresistivities (panel b) of monolayer graphene as a function of gate voltage, $V_g$, from $-80$ to $80$ V. The simulations were performed at a temperature $T = 1.6$ K, magnetic field $B = 9$ T, gyromagnetic factor $g^* = 2$, and effective mass $m^* $= 0.0124 $m_0$. The Gaussian Landau-level broadening was set to $\Gamma = 0.06$ eV, and the electron relaxation time to $\tau = 1$ ps. In panel (a), the integer numbers above the reference lines indicate the corresponding Hall plateaux.}
    \label{fig:placeholder1}
\end{figure}

\begin{figure}
    \centering
    \includegraphics[width=0.5\linewidth]{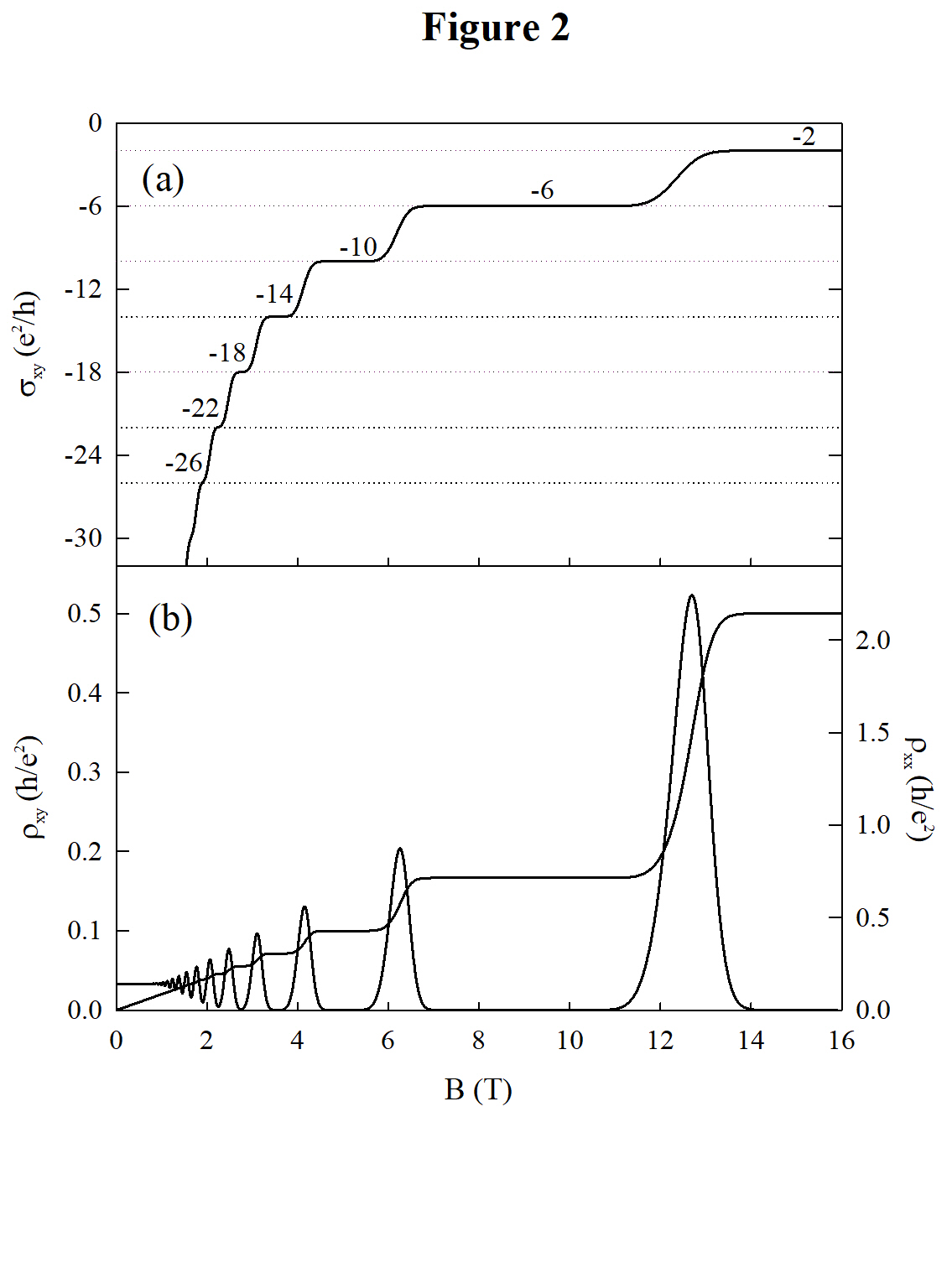}
    \caption{Simulation of the Hall magnetoconductivity (panel a) and diagonal magnetoresistivity (panel b) of monolayer graphene as a function of magnetic field, $B$. The simulation parameters are the same as in Figure 1, except that the electron density is $n_0 = 6 \times 10^{15} , \mathrm{m}^{-2}$. In panel (a), the integer numbers above the reference lines indicate the corresponding Hall plateaux.}
    \label{fig:placeholder2}
\end{figure}

\begin{figure}
    \centering
    \includegraphics[width=0.5\linewidth]{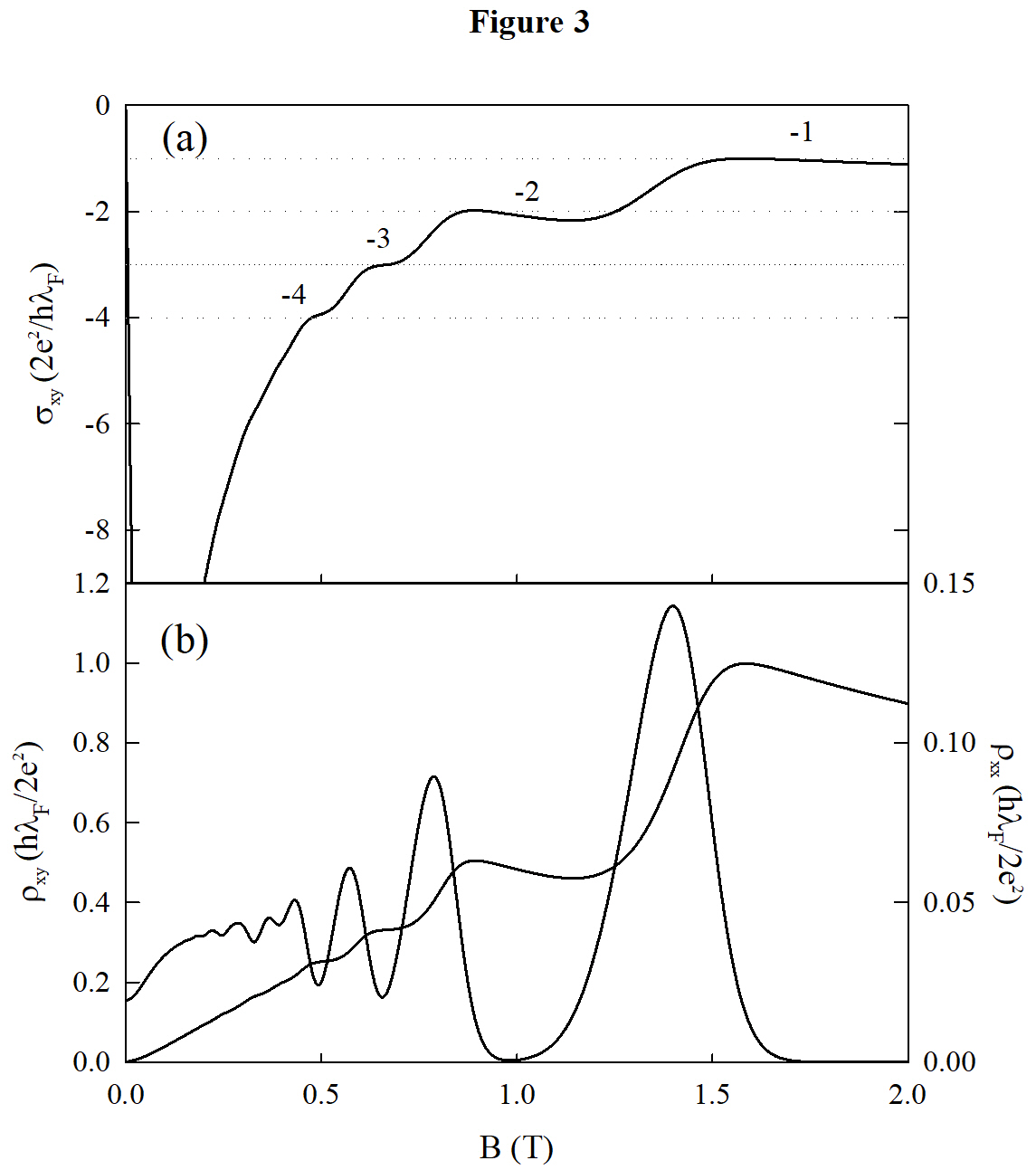}
    \caption{Simulation of the Hall magnetoconductivity (panel a) and the diagonal and Hall magnetoresistivities (panel b) of a 3D semimetal as a function of magnetic field, $B$. The simulations were performed with a carrier density $n_0 =10^{22}$ $m^{-3}$, at a temperature $T = 0.06$ K, effective mass $m^* = 0.04m_0$, gyromagnetic factor $g^* = 27$, and cyclotron frequency correction factor $\alpha = 2/3$. The Gaussian Landau-level broadening was set to $\Gamma = 0.002$ eV, and the electron relaxation time to $\tau = 1$ ps. In panel (a), the integer numbers above the reference lines indicate the corresponding Hall plateaux, while panel (b) shows the diagonal resistivity (Shubnikov–de Haas oscillations).}
    \label{fig:placeholder3}
\end{figure}

\begin{figure}
    \centering
    \includegraphics[width=0.5\linewidth]{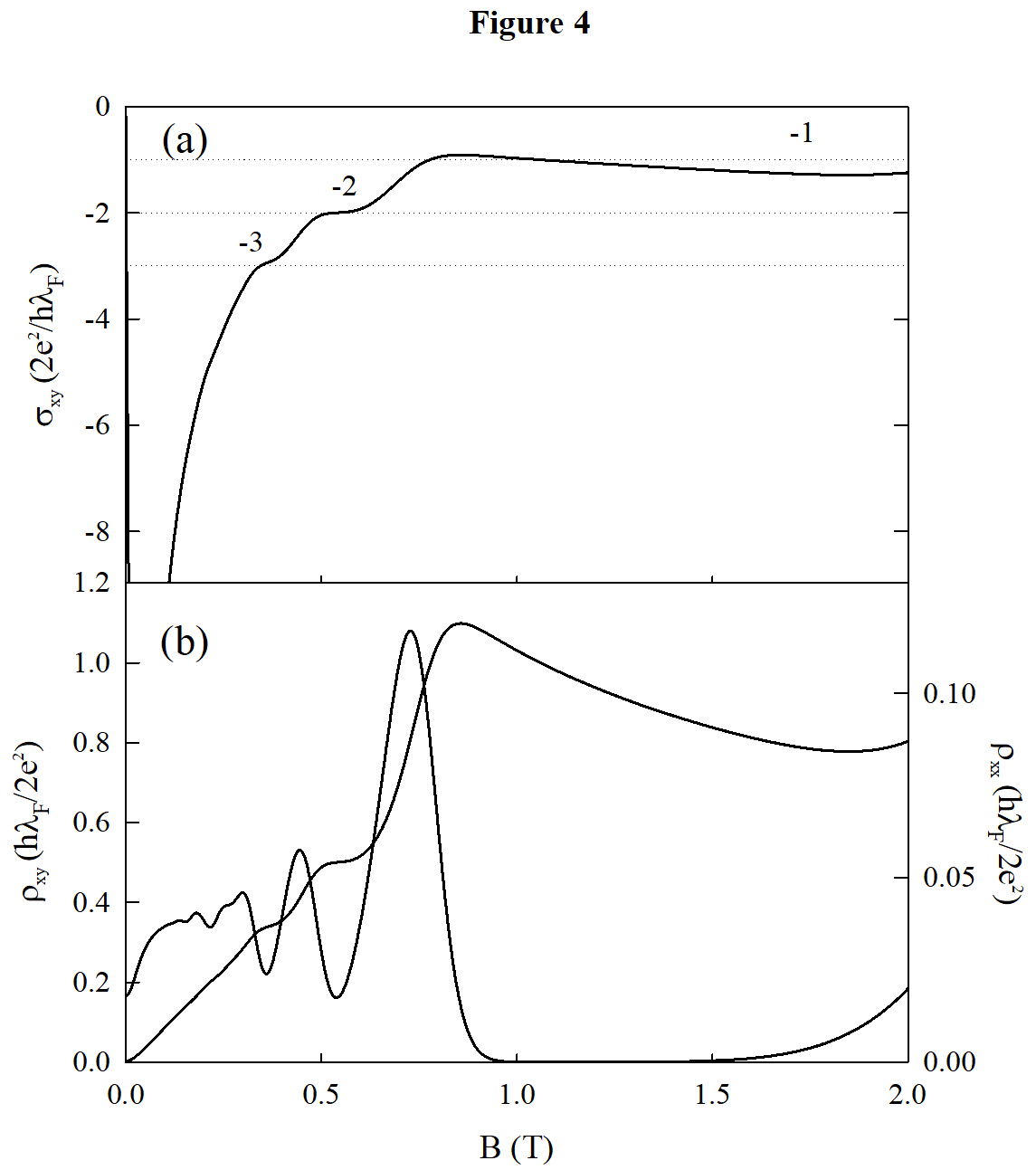}
    \caption{Simulation of the Hall magnetoconductivity (panel a) and the diagonal and Hall magnetoresistivities (panel b) of a 3D semimetal as a function of magnetic field. Simulations were performed with a carrier density $n_0 = 4 \times 10^{21} \mathrm{m}^{-3}$, temperature $T = 0.07$ K, effective mass $m^* = 0.03 m_0$, gyromagnetic factor $g^* = 30$, and cyclotron frequency correction factor $\alpha = 2/3$. Gaussian Landau-level broadening $\Gamma = 0.001$ eV and electron relaxation time $\tau = 6$ ps were used. In panel (a), integers above the reference lines indicate Hall plateaux; panel (b) shows the diagonal resistivity (Shubnikov–de Haas oscillations).}
    \label{fig:placeholder4}
\end{figure}

\end{document}